\documentclass[aps,prb,twocolumn,preprintnumbers,amsmath,amssymb]{revtex4}
\usepackage{graphicx}
\usepackage{float}
\usepackage{xcolor}
\begin{document}
\title{Linear and isotropic magnetoresistance of Co$_{1-x}$Fe$_x$Si at x=0.2; 0.4; 0.65.}

\author{A.~E.~Petrova}
\affiliation{P.~N.~Lebedev Physical Institute, Leninsky pr., 53, 119991 Moscow, Russia}
\author{S.~Yu.~Gavrilkin}
\affiliation{P.~N.~Lebedev Physical Institute, Leninsky pr., 53, 119991 Moscow, Russia}
\author{S.~S.~Khasanov}
\affiliation{Institute of Solid State Physics RAS, Chernogolovka, Moscow District,142432 Russia}
\author{V.~A.~Stepanov}
\affiliation{P.~N.~Lebedev Physical Institute, Leninsky pr., 53, 119991 Moscow, Russia}
\author{Dirk Menzel}
\affiliation{Institut f\"{u}r Physik der Kondensierten Materie, Technische Universit\"{a}t Braunschweig, D-38106 Braunschweig, Germany}
\author{S.~M.~Stishov}
\email{stishovsm@lebedev.ru}
\affiliation{P. N. Lebedev Physical Institute, Leninsky pr., 53, 119991 Moscow, Russia}

\begin{abstract}
We studied the magnetoresistance (MR) of well-characterized samples of Co$_{1-x}$Fe$_x$Si at x=0.2, 0.4, and 0.65 at temperatures between 1.8 and 100~K and magnetic fields of 9~T. The quasilinear dependence of MR on the magnetic field at low temperatures and the practically isotropic properties of MR in these compounds are tentatively attributed to the specifics of Weyl electron spectra and general disorder of the materials. 
\end{abstract}

\maketitle

\section{Introduction}
\begin{figure}[htb]
\includegraphics[width=40mm]{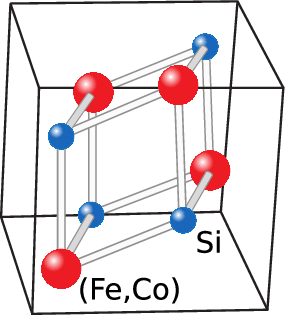}
\caption{\label{fig1} Chiral crystal structure of binary compounds B20.}
\end{figure}

\begin{figure}[htb]
\includegraphics[width=80mm]{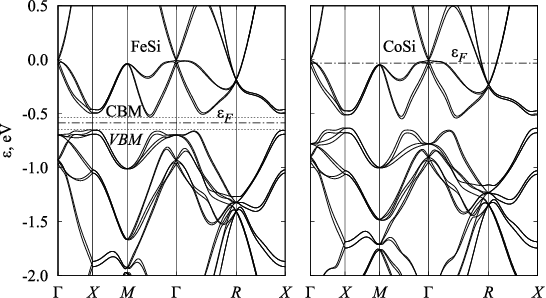}
\caption{\label{fig2} Band structure of FeSi, and CoSi with a B20
(P2$_1$3) crystal structure. Energy is measured relative to the fourfold
intersection of the bands at point $\Gamma$. Dashed-dotted line shows position of the Fermi level~\cite{Psh}.}
\end{figure}

\begin{figure}[htb]
\includegraphics[width=80mm]{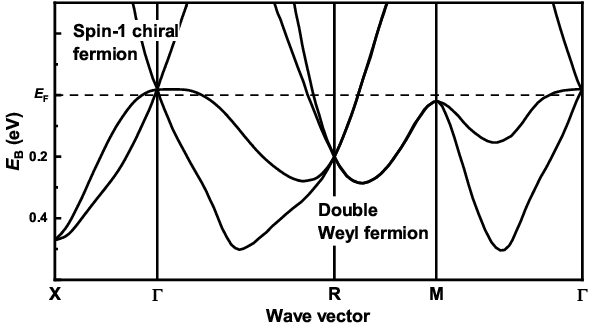}
\caption{\label{fig3} Band structure of CoSi near $E_F$ along high-symmetry directions
of the Brillouin zone calculated~\cite{Psh} without taking into account spin-orbit interaction~\cite{Tak}.}
\end{figure}

\begin{figure}[htb]
\includegraphics[width=80mm]{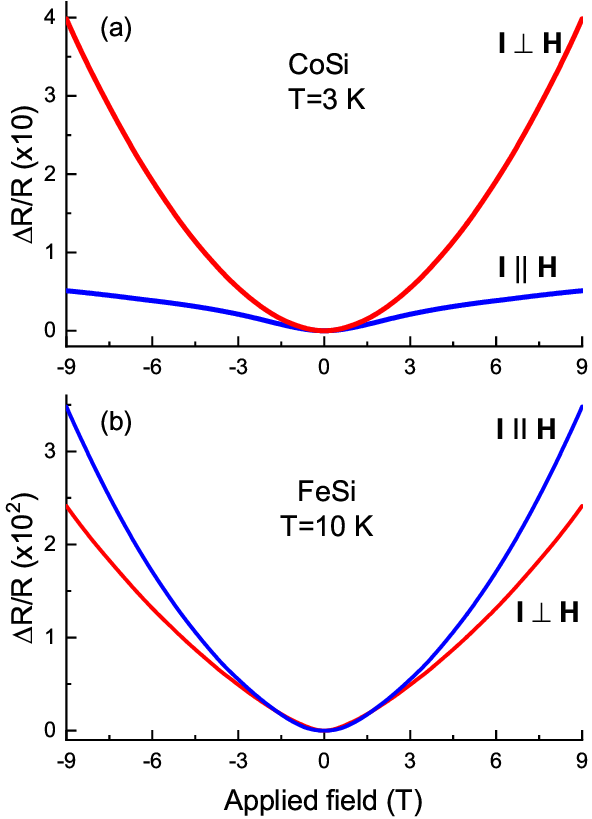}
\caption{\label{fig4} Magnetoresistance of the parent compounds CoSi (a)~\cite{Pet} and FeSi (b)~\cite{Pet2}. Note nontrivial behavior of MR in semiconductive FeSi.}
\end{figure}

Binary compounds of transition metals that crystallize in the chiral B20 crystal structure (Fig.\ref{fig1}) demonstrate a number of remarkable physical properties. In particular, their electron spectra reveal topological features known as Weyl nodes (Fig.~\ref{fig2},~\ref{fig3})~\cite{Psh,Tak}. These features provide the basis for forming a Weyl semimetal with two Weyl cones separated in momentum space where the conduction and valence bands touch at discrete points with a linear dispersion relation. This guarantees that any Weyl fermions with opposite chiralities are separated in momentum space as well. Thus, parallel magnetic and electric fields can move electrons between Weyl cones of opposite chirality, giving rise to negative longitudinal magnetoresistance. This situation is called the chiral anomaly. Of all the compounds with a chiral B20 crystal structure, cobalt silicide (CoSi) seemed to be an appropriate candidate for observing the effects of the chiral anomaly. Indeed, CoSi is a conductive material with fermionic excitations within the Fermi level.
The first attempts to observe negative magnetoresistance in CoSi samples were unsuccessful~\cite{Pet}, probably due to the current jetting effect~\citep{Pip}. Later developments involved performing MR measurements using microsamples prepared by lithographic techniques, which resulted in the discovery of negative longitudinal magnetoresistance (NLMR)~\cite{Sch,Bal}. Note that the NLMR was observed in a Co$_{1-x}$Fe$_x$Si system at x=0.02-0.06 in work~\cite{Sch}. The NLMR is not observed at values of $x>0.06$. However, surprising isotropic behavior~\cite{Sch} of the MR is evident at $x>0.1$. Almost isotropic and linear MR in the sample Co$_{15}$Fe$_{85}$Si is found in Ref.~\cite{Man}. Nearly the isotropic MR is also seen in Ref.~\cite{SX} even at 30~K. Typically, one would expect a quadratic increase in resistance with the magnetic field in the case of a transverse configuration and a slight response in resistance when the magnetic field is applied parallel to the current (see Fig.~\ref{fig4}, for example). This kind of MR behavior obviously leads to significant anisotropy. On the other hand, as shown in Fig.~\ref{fig4}b, the situation is not so simple and there is no satisfactory theoretical understanding the observed experimental data. With all these in mind, we decided to take a closer look at the state of things by studying the MR of Co$_{1-x}$Fe$_x$Si at x=0.2, 0.4, and 0.65.

\section{Experimental} 
Sample preparation was performed by melting 99.98 percent of Fe, 99.9 percent Co, and Si of semi conductive quality in a single arc oven under argon atmosphere. From this melt single crystals with a diameter of approximately 5 mm were grown using the tri-arc Czochralski technique. 

The chemical composition of the samples was determined by electron probe microanalysis. The lattice parameters were obtained from the Rietveld analysis of the X-ray powder diffraction. The samples selected for magnetoresistance measurements have dimensions of $2.5 \times 0.9 \times 0.6~ mm^{3}$. 
The heat capacity of the samples were measured using the Quantum Design physical property measurement system with the heat capacity module and He-3 refrigerator. Resistivity was measured using the standard four-terminal scheme with gold wires bonded to the sample by the silver paste as electrical contacts. Magnetic susceptibilities were measured using a Quantum Design magnetic property measurement system. 

Magnetoresistance measurements were performed in two orientations with the current parallel $I\| H$ and perpendicular $I\bot H$ to the magnetic field. During the measurements, the sample was cooled from 300~K to 1.8~K in steps (see Fig.~\ref{fig5}, \ref{fig6}). At each step the magnetic field was applied from 0 to 9~T and the resistance was measured in the $I\bot H$ configuration, then the measurement was continued in the $I\|H$ configuration when the magnetic field was decreased from 9~T to 0. The physical characteristics of the samples obtained are illustrated in the Appendix.

\subsection{Experimental results}
\begin{figure}[htb]
\includegraphics[width=80mm]{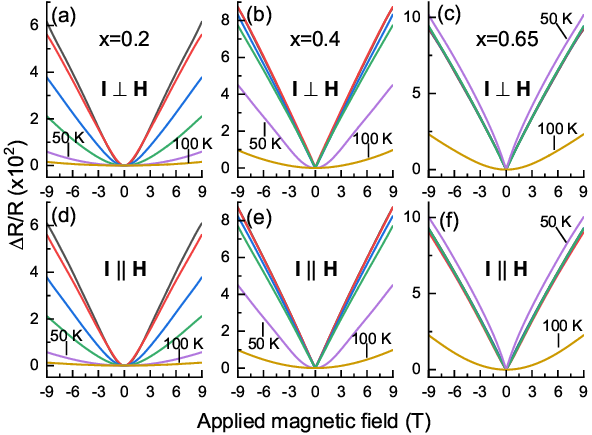}
\caption{\label{fig5} Longitudinal and transverse magnetoresistance of Co$_{1-x}$ Fe$_{x}$Si as functions of magnetic field and temperature.}
\end{figure}

\begin{figure}[htb]
\includegraphics[width=80mm]{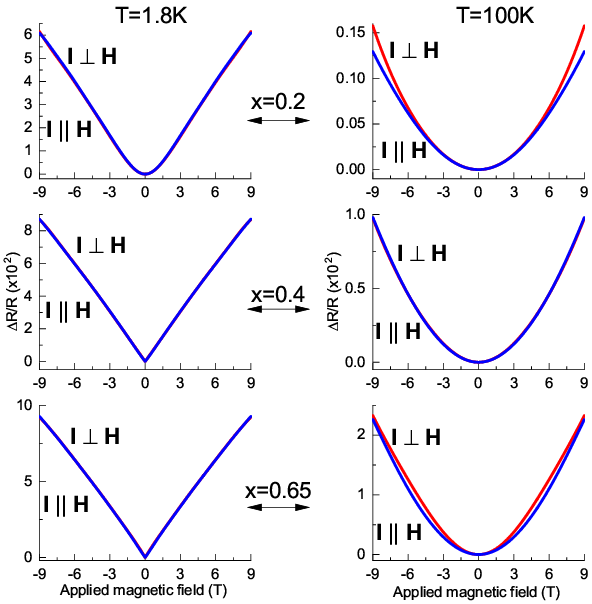}
\caption{\label{fig6} Isotropic magnetoresistance of Co$_{1-x}$Fe$_{x}$Si. MR curves for $I\| H$ and $I\bot H$ in the left panel and at x=0.4 of right panel can not be distinguished in our measurements. Isotropy disappears at 100~K for the sample with x=0.2, whereas it still holds for the samples with x=0.4 and 0.65. The mismatch between the two curves in the last sample is due to noisy experimental data. It is significant that MR of the sample with x=0.2 does not show a quasilinear behavior at low temperature (see also Fig.~\ref{fig5}). }
\end{figure}

All longitudinal and transverse magnetoresistance data as functions of magnetic field and temperature are displayed in Fig.~\ref{fig5}. As clearly seen in the case of two materials with x=0.4 and 0.65, MR(H) is a quasilinear function of the magnetic field in both configurations $I\| H$ and $I\bot H$ at low temperatures. This type of behavior changes with temperature and the MR(H) dependence acquires a customary form. For example, approximating MR at 100~K of all three substances by a power expression $\Delta \rho/\rho \sim H^{n}$ one finds usual values of the exponent n varies from 1.8 to 2.           
The low temperature (1.8~K) behavior of the material with x=0.2 is slightly different from others two. At low magnetic fields the dependence MR(H) of this substance shows some curvature, which straitens up at high magnetic fields and became ideally linear with n=1, if one would use again the same power expression as above. Incidentally an application of aforementioned power expression to the other three quasilinear lines yields at T=1.8~K values n ranging from 0.93 to 0.96. It should be noted  here that the sample with x=0.2 indeed is different from the rest two, it behaves as a metal below 50~K with respect to electrical conductivity and doesn't show any indications for a spin ordered structure.
The MR anisotropy test was performed by plotting data of both configurations together, as shown in Fig.~\ref{fig6}. As can be seen, there is surprising, almost perfect isotropy of MR at T=1.8~K, which disappears at 100 K~for the sample with x=0.2. However, it still holds for the samples with x=0.4 and 0.65. The mismatch between the two curves in the last sample is due to noisy experimental data.

\section{Discussion}
As was demonstrated in the previous section, two remarkable features characterize the magnetoresistance of Co$_{1-x}$Fe$_x$Si samples at x=0.2, 0.4, and 0.65. These features are the linear dependence of both the transverse and longitudinal magnetoresistance (MR) on the magnetic field and an almost perfect isotropic property of the MR, meaning it is independent of the direction of the magnetic field relative to the current. The linear MR has been found in a number of substances, and its nature has been investigated (see for instance~\cite{Ab,Par, Hu, Son}). Most cases that have been considered, are somehow connected to the disorder or inhomogeneity in the materials. However, for so-called "quantum linear magnetoresistance" probably enough to have gapless spectrum with a linear momentum dispersion~\cite{Ab}. The quasilinear MR in the Weyl semimetals Co$_{1-x}$Fe$_{x}$Si observed in the current work (Fig.~\ref{fig5}) most probably has this kind of quantum origin. However, it can be clearly seen that MR of the sample with x=0.2 demonstrate almost normal, close - to - quadratic behavior at low temperature. Note that samples with x = 0.2 show no signs of magnetic ordering, while samples with x = 0.4 and 0.65 are magnetically ordered at low temperatures (see Appendix). This probably implies that a spin polarization plays an essential role in the MR behavior. As expected, MR turns to the classical quadratic field dependence $\approx H^2$ at elevated temperature (Fig.~\ref{fig6}). On the other hand the isotropic features of the MR in Co$_{1-x}$Fe$_{x}$Si (Fig.~\ref{fig6}) is certainly have a classical character showing up even at 100~K. As mentioned in Ref.~\cite{Hu}, when gross inhomogeneities exist in a semiconductor it is possible to create distorted current paths misaligned with the driving voltage, and mix in the off-diagonal components of the magnetoresistivity tensor. As a result there will be no difference between longitudinal and transverse directions of magnetic field. Our samples of Co$_{1-x}$Fe$_x$Si are disordered solid solutions and that is probably a key reason for the observed MR isotropy. In this connection see Ref.~\cite{How} where MR isotropy was discovered in the amorphous material.
\section{Conclusion}
Two remarkable features characterize the magnetoresistance of Co$_{1-x}$Fe$_x$Si samples at x = 0.2, 0.4, and 0.65 that are a linear dependence of both the transverse and longitudinal magnetoresistance (MR) on the magnetic field and an almost perfect isotropic property of the MR. These properties are tentatively attributed to the specifics of Weyl electron spectra and general disorder of the materials. However, it should be noted that a spin polarization likely has a significant impact  on the MR  behavior in Co$_{1-x}$Fe$_x$Si samples. 
\section{acknowledgment}
We would like to express our sincere gratitude to Dr. K.I. Kugel for reading the manuscript and discussion and to Dr. E.G. Nikolaev for the technical assistance.

\section{Appendix} 
\subsection*{Sample characterizations}
\begin{figure}[htb]
\includegraphics[width=60mm]{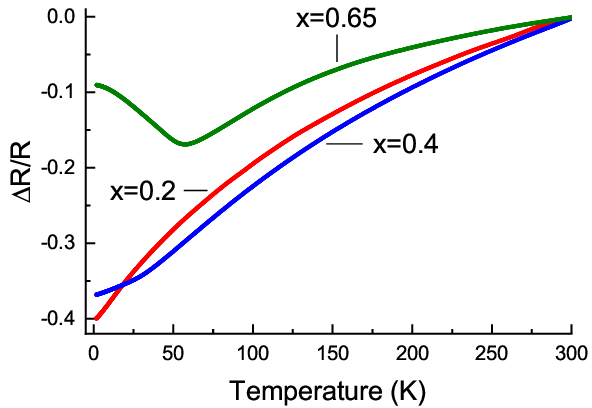}
\caption{\label{fig7} Reduced electrical resistivity as function of temperature for Co$_{1-x}$Fe${_x}$Si at x=0.2; 0.4; 0.65. The Fe-rich sample with x=0.65 indicates a semiconducting behavior at $T<50K$.}
\end{figure}

\begin{figure}[htb]
\includegraphics[width=60mm]{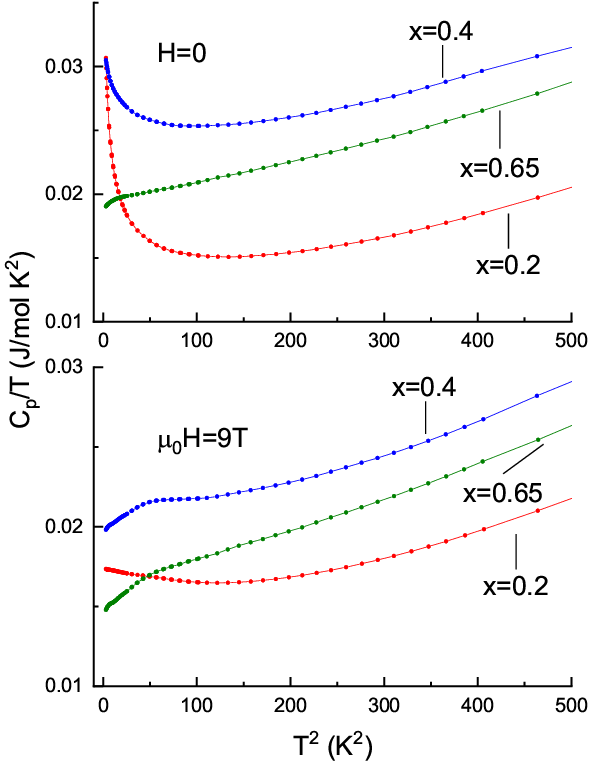}
\caption{\label{fig8} Ratio $C_p/T$ as function of temperature for Co$_{1-x}$Fe$_{x}$Si at x=0.2; 0.4; 0.65. Is seen a tendency to quantum critical behavior at $T\rightarrow 0$ for x=0.2 and x=0.4.}
\end{figure}

\begin{figure}[H]
\includegraphics[width=60mm]{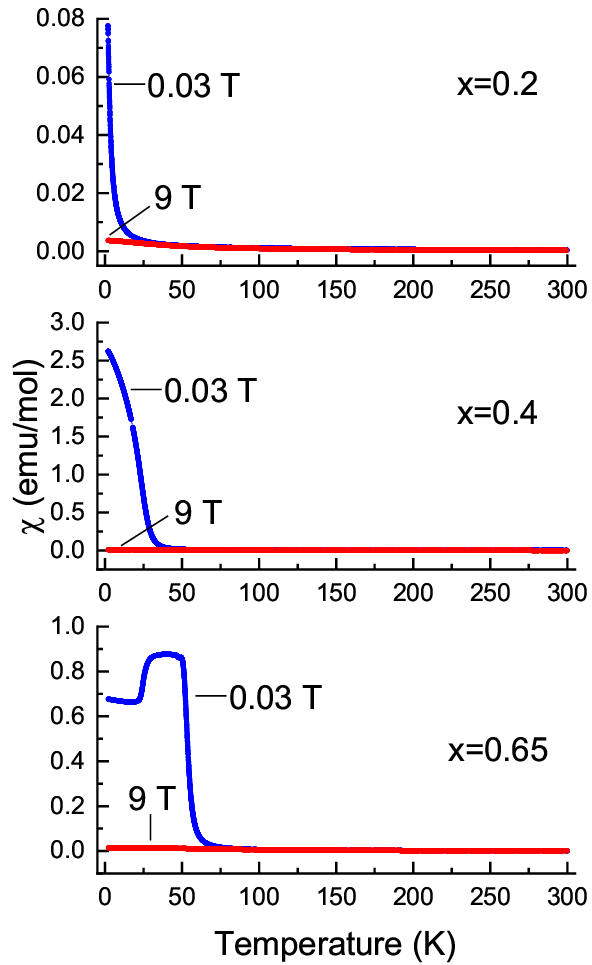}
\caption{\label{fig9} Magnetic Susceptibility of Co$_ {1-x}$Fe$_ {x}$Si as functions of temperature. The data Co$_{0.35}$Fe$_{O.65}$ Si and Co$_{0.6}$Fe$_{O.4}$ possibly indicate an existence of skryrmions, which is typical of (Fe,Co)Si system~\cite{Bei}. These features are not seen in the heat capacity curve at H=0 and $\mu_0$H=9~T as expected (Fig.~\ref{fig8}).}
\end{figure}

\begin{figure}[H]
\includegraphics[width=60mm]{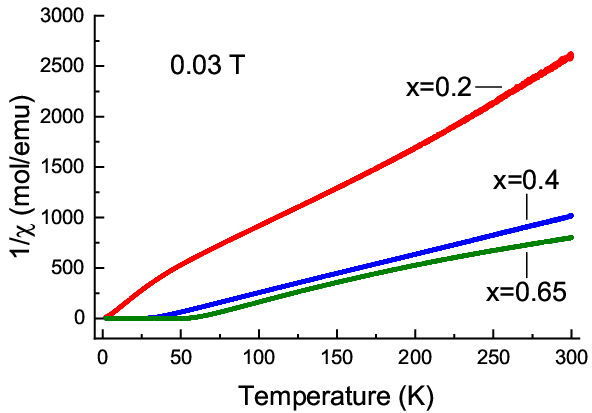}
\caption{\label{fig10} The inverse susceptibility data indicate once again that Co$_{0.8}$Fe$_{O.2}$ is a simple paramagnetic, whereas Co$_{1-x}$Fe$_{x}$ at x=0.4 and 0.65 show complicated behavior as suggested in Fig.~\ref{fig9}.}
\end{figure}

The lattice parameters of Co$_{1-x}$Fe$_{x}$Si are generally agree with the results of Ref.~\cite{Sch} at low x, though a fulfillment of the Vegard rule is questionable. The resistivity data are shown in Fig.~\ref{fig7}. It is look like Fe-rich sample Co$_{0.35}$Fe$_{0.65}$Si demonstrates a semiconducting behavior at low temperatures. The heat capacity data are demonstrated in Fig ~\ref{fig8}. The diverging character of $C_p/T$ of samples with x=0.2 and 0.4 is obvious and indicates a tendency to quantum critical behavior, which suppressed by magnetic field application.
The magnetic susceptibility of the samples are illustrated in Figs.~\ref{fig9}-\ref{fig10}. As is seen the sample with x=0.2 reveals almost typical paramagnetic behavior, whereas the samples with x=0.4 and 0.65 probably indicate an existence of skryrmions, which is typical of (Fe,Co)Si system~\cite{Bei}. These features are not seen in the heat capacity curves at H=0 and $\mu_0$H=9~T as expected (Fig.\ref{fig8}). These data agree with the existence of helical phase transition reported in Ref.~\cite{On}.


\end{document}